\documentclass[%
 reprint,
nofootinbib,
 amsmath,amssymb,
 aps,
]{revtex4-2}

\usepackage{graphicx}
\usepackage{dcolumn}
\usepackage{bm}
\usepackage{physics}
\usepackage[version=4]{mhchem}
\usepackage{siunitx}
\usepackage[hidelinks]{hyperref}

\begin{document}


\title{Nonequilibrium Green's Function simulation of \ce{Cu2O} photocathodes for photoelectrochemical hydrogen production}
\let\thefootnote\relax\footnotetext{© 2023 American Physical Society}

\author{Lassi Hällström}
\email{lassi.hallstrom@aalto.fi}
\author{Ilkka Tittonen}

\affiliation{
Department of Electronics and Nanoengineering, Aalto University, Finland
}%




\date{\today}

\begin{abstract}
 In this work we present a simulation of the semiconductor electrodes of photoelectrochemical (PEC) water splitting cells based on the nonequilibrium Green's function (NEGF) formalism. While the performance of simple PEC cells can be adequately explained with semi-classical drift-diffusion theory, the increasing interest towards thin film cells and nanostructures in general requires theoretical treatment that can capture the quantum phenomena influencing the charge carrier dynamics in these devices. Specifically, we study a p-type \ce{Cu2O} electrode and examine the influence of the bias voltage, reaction kinetics and the thickness of the \ce{Cu2O} layer on the generated photocurrent. The NEGF equations are solved in a self–consistent manner with the electrostatic potential from Poisson’s equation, sunlight induced photon scattering and the chemical overpotential required to drive the water splitting reaction. We show that the NEGF simulation accurately reproduces experimental results from both voltammetry and impedance spectroscopy measurements, while providing an energy resolved solution of the charge carrier densities and corresponding currents inside the semiconductor electrode at nanoscale.

\end{abstract}

\maketitle


\section{\label{sec:Introduction}Introduction:\protect}

Photoelectrochemical (PEC) water splitting offers substantial potential for carbon free hydrogen generation. An accurate understanding of the underlying physics is crucial for educated design and optimization of future cell designs. PEC cells have been successfully modeled on multiple different levels ranging from full device-level models to ab-initio models of the chemical catalyst operation. The operation of the semiconductor electrode, the key light-absorbing component of the cell, has been simulated using the well established drift-diffusion theory.~\cite{cavassilas2014modeling,hallstrom2021computational} However, the drift-diffusion model is limited by its underlying assumptions as the electrode designs move into the nanoscale, with more complex materials and thin film structures. 
In nanoscale electronics, nonequilibrium Green's function (NEGF) approach has been widely applied in transistor design~\cite{ren2003nanomos,dastjerdy20113d,pal2012electron}, and in general nanoscale conductance~\cite{guo2004toward} and more recently in solar cell modeling~\cite{aeberhard2011theory,aeberhard2012nonequilibrium,aeberhard2014quantum,cavassilas2014modeling,aeberhard2019challenges}. It offers an excellent compromise between the limiting assumptions included in classical transport equations and computationally heavy atomic level ab initio quantum chemical simulations, while providing an energy resolved picture of the charge dynamics in the semiconductor. The ability to resolve the energy spectrum of the generated current is especially critical for simulating electrochemical water splitting devices, as the efficiency depends on its ability to drive an endothermic chemical reaction, and the reaction rate is directly tied to the energy that the charge carriers have available.

Various metal oxides such as \ce{TiO2}, \ce{FeO2}, \ce{Cu2O}, have been extensively studied due to their relatively high efficiency and low cost giving them potential to scale up to industrial scale hydrogen production.~\cite{wick2015photovoltaic,azevedo2014stability,singh2021hydrogen} As these materials are commonly used as thin films with thicknesses below \SI{1}{\micro m}, all the way down to just a few nanometers\cite{paracchino2012ultrathin,le2010controlling,yang2020recent}, a microscopic description of the carrier dynamics allows resolving the nonequilibrium behaviour of the electrode at nanoscale.
In this work we apply NEGF formalism to simulate the operation of the light capturing semiconductor electrode of a thin film device. We consider a \SI{50}{nm} thick \ce{Cu2O} electrode acting as a photocathode driving the hydrogen evolution reaction in a water splitting PEC cell. We demonstrate that the NEGF formalism with proper treatment of the boundary conditions can be applied to simulating PEC systems. The simulation is shown to reproduce the typical current-voltage behaviour as well as the Mott-Schottky relationship between the measured voltage and the surface capacitance that is observed in experimental studies. Furthermore, the impact of the semiconductor thickness and the reaction kinetics on the generated photocurrent is examined.




\section{\label{sec:Computational model}Computational model}
\subsection{The Green's functions}
\begin{figure}
\includegraphics[width=\columnwidth]{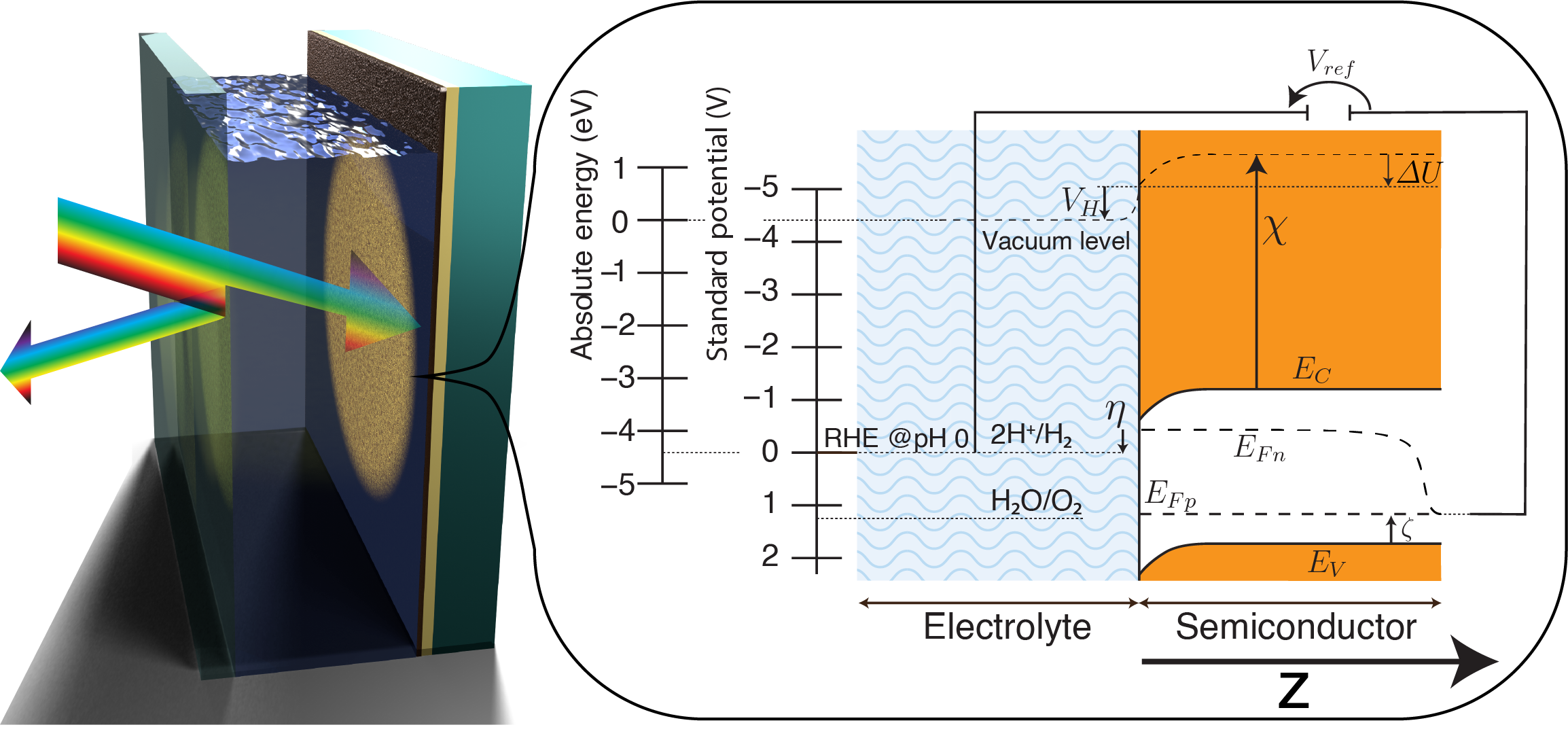}
\caption{\label{fig:energydiagram}Schematic of a photoelectrochemical cell and the energy band diagram of the water splitting photocathode. Sunlight enters the cell through a glass window and the liquid  electrolyte, with some light lost to reflection and absorption. The reference potential is the hydrogen evolution reaction potential as measured by a RHE. The simulation controls $V_{ref}$, and solves the quasi Fermi levels for both carriers and the band edge shape. $V_H$ is the Helmholtz potential in the electric double layer on the electrolyte side of the interface, $\eta$ the overpotential for the hydrogen evolution, and $\zeta$ the distance of the Fermi level from the valence band edge at equilibrium.}
\end{figure}

\begin{table}[b]
\caption{\label{tab:table1}%
Material parameters used for the simulation
}
\begin{ruledtabular}
\begin{tabular}{lll}
\textrm{Symbol}&
\textrm{Parameter}&
\textrm{Value}\\
\colrule
$d$ & sample thickness & \SI{50}{nm}\\
$E_g$ & bandgap & \SI{2.17}{eV} \\
$\hbar\omega$ & optical phonon energy & \SI{88.4}{meV}\\
$j_0$ & exchange current density & \SI{5e-6}{A m^2}\\
$m_e^*$ & electron effective mass & \SI{0.98}{}\\
$m_h^*$ & hole effective mass & \SI{0.58}{}\\
$N_a$ & doping density & 5e17 cm\textsuperscript{-3}\\
$V_{fb}$ & flat band potential & \SI{1.15}{V} vs. RHE\\
$\varepsilon_r$ & relative permittivity & \SI{6.0}{}\\
$\alpha$ & photon absorption coefficient & \SI{1.2e5}{cm^{-1}}\footnote{Fitted to the absorption from the computation of the photon scattering self-energy.}\\
\end{tabular}
\end{ruledtabular}
\label{tb:material}
\end{table}
The energy band diagram of the PEC cell is shown in Fig.~\ref{fig:energydiagram}. The controlled voltage $V_{ref}$ is defined against the reversible hydrogen electrode (RHE), which directly gives the reaction potential of the hydrogen evolution reaction. At the semiconductor electrolyte interface (SEI) the band edge potentials are fixed vs. the reference potential by the electron affinity $\chi$ and the Helmholtz potential $V_H$. In practice however, the Helmholtz potential is generally unknown and the band edge positions are determined by measuring the flat band potential $V_{fb}$ of the electrode.~\cite{park2019optimal,rajeshwar2018copper} The flat band potential defines the value of $V_{ref}$ at which the electrostatic potential drop across the semiconductor is zero, and therefore the conduction and valence bands are flat. Fixing the value of the flat band potential therefore also defines the band edge potentials at the SEI.

In this work, we consider the two-band effective mass approximation for simulating the band structure of the \ce{Cu2O} electrode. The material parameters are shown in Table~\ref{tb:material}. The non-interacting Hamiltonian describing the carrier dynamics in the semiconductor electrode is given by 

\begin{align}
H_{b,k_{\parallel}} = -\frac{\hbar^2}{2 m_{b}^{*}} \frac{d^2}{dz^2} - eU(z) + \frac{\hbar^2 k_\parallel^2}{2m_b^*} + E_b,
\end{align}
where  $m_{b}^{*}$ is the effective mass of the charge carrier in band $b$, $z$ is the position in real space in the normal direction of the semiconductor surface, $e$ the elementary charge, $U$ the electrostatic potential, $ k_\parallel$ the transverse momentum and $E_b$ the band edge potential at the SEI. In this formalism the effective mass of the valence band holes is negative. In the tight-biding approximation only the neighbouring spatial elements are coupled, and the Hamiltonian has a tridiagonal finite difference matrix representation

\begin{align}
H_{b,k_\parallel} = \mqty[T_{b,k_\parallel} & -t_b & \cdots & 0 & 0\\ 
           -t_b & T_{b,k_\parallel} & \cdots & 0 & 0\\ 
            \vdots & \vdots & \ddots & \vdots & \vdots\\ 
            0 & 0 & \cdots & T_{b,k_\parallel} & -t_b\\ 
            0 & 0 & \cdots & -t_b & T_{b,k_\parallel}]
\end{align}
where
\begin{align}
    t_b =&\frac{\hbar^2}{2 m_b^* \Delta z^2}\\
    T_{b,k_\parallel} =&-eU(z) + \frac{\hbar^2 k_\parallel^2}{2m_b^*} + E_b + 2t_b.
\end{align}
The Green's functions describing the steady-state charge carrier distribution are obtained from the Dyson equation
\begin{align}
\label{eq:Dyson}
G^r_{b,k_\parallel} &= [EI - H_{b,k_\parallel} - \Sigma^r_{b,k_\parallel}]^{-1}
\end{align}
where $G^r_{b,k_\parallel}$ is the retarded Green's function, $E$ is the energy, $I$ is the identity matrix, and $\Sigma^r_{b,k_\parallel}$ the retarded self energy term describing the scattering processes present in the simulated system and the boundary conditions from the electrical contacts. The lesser and greater components of the Green's function can the be solved form the Keldysh equation
\begin{align}
\label{eq:Ret}
G_{b,k_\parallel}^\lessgtr &= G_{b,k_\parallel}^r \Sigma^{\lessgtr}_{b,k_\parallel} G_{b,k_\parallel}^{r\dagger},
\end{align}
 where $\Sigma^{\lessgtr}_{b,k_\parallel}$ are the lesser and greater self energies.

\subsection{Scattering processes}
For scattering processes, we apply the local self-energy approximation. This has the consequence that the self energy terms in Eq.~\ref{eq:Dyson} are diagonal matrices, which preserves the tridiagonal form of the right hand side and offers considerable savings in the computation time needed to solve the linear systems. The primary scattering mechanism to consider in a PEC cell is the photon scattering from the incident solar radiation. As solar radiation is a wideband light source, the photon scattering self energies must be integrated over the energy range of the incoming photons, limited by the bandgap of the semiconductor and atmospheric absorption of uv radiation.
\begin{align}
\label{eq:photonSEc}
\Sigma^{\lessgtr}_{\nu(c)}(E) = i \mathfrak{Im} \qty(\int d(h\nu) M^2 \rho_\nu G^{\lessgtr}_{\nu(v)}(E-h\nu)) \\
\label{eq:photonSEv}
\Sigma^{\lessgtr}_{\nu(v)}(E) = i \mathfrak{Im} \qty(\int d(h\nu) M^2 \rho_\nu G^{\lessgtr}_{\nu(c)}(E-h\nu))
\end{align}
where the subscripts $c$ and $v$ stand for conduction and valence band, $\nu$ is the photon frequency,
\begin{align}
\label{eq:photonelectronM}
M = \frac{e\hbar^2}{2m^*} \sqrt{\frac{1}{\varepsilon_r\varepsilon_0 h \nu}}
\end{align}
is the photon-electron coupling factor and 
\begin{align}
\label{eq:beerlambert}
\rho_\nu = \alpha e^{-\alpha z}  \frac{\Phi_\nu}{c}
\end{align}
the photon density. Here, $\alpha$ is the absorption constant for the semiconductor material and $\Phi$ the incident AM1.5G spectral photon flux attenuated by the glass window and the water based electrolyte found in typical experimental PEC cells. The optical attenuation is computed using a transfer matrix method accounting for both reflection and absorption losses.~\cite{hallstrom2021computational,katsidis2002general} With \SI{2}{mm} of glass and \SI{2}{cm} of water, the transmission efficiency for photons with energy above the bandgap of the \ce{Cu2O} electrode studied in this work is 90\%. After reaching the semiconductor, the photon flux is absorbed by the semiconductor material via the scattering process defined by the self energies in Eqs.~\ref{eq:photonSEc} and ~\ref{eq:photonSEv}. The optical absorption coefficient $\alpha$ is matched to the simulation so that the absorption rate implied by Eq.~\ref{eq:beerlambert} matches with the carrier generation rate found in the converged solution. It is noteworthy that even though the local approximation of the scattering self energies is known to underestimate the scattering rates~\cite{kubis2011assessment,aeberhard2014quantum,cavassilas2016local,aeberhard2020microscopic}, the value of the absorption coefficient in Table~\ref{tb:material} resulting from the absorption model in Eqs.~\ref{eq:photonSEc} and \ref{eq:photonSEv} is on the high end of reported experimental values for \ce{Cu2O}.~\cite{rakhshani1987optical,malerba2011absorption} Nevertheless, for precise quantitative results of the scattering rates the nonlocal nature of the scattering processes should be accounted for either directly or via a compensation factor like described in~\cite{sarangapani2019band}.
For phonon scattering we include a rudimentary model of polar optical phonon scattering with the assumption of a constant phonon energy of \SI{88.4}{meV}~\cite{jo2022experimental}. The phonon scattering self energies can be calculated as

\begin{align}
\Sigma^{\lessgtr}_{\nu(b)}(k_\parallel,E) = \frac{\gamma \pi}{(2\pi)^3} \int dl_\parallel V_{p}(\abs{k_\parallel - l_\parallel})\nonumber\\ 
\qty(N_{LO}G^{\lessgtr}(l_\parallel,E \mp \hbar\omega) + (N_{LO}+1)G^{\lessgtr}(l_\parallel,E \pm \hbar\omega)),
\end{align}
where $\gamma$ is the Frölich coupling constant~\cite{frohlich1952interaction}
\begin{align}
\gamma = e^2\frac{\hbar\omega}{2\varepsilon_0}\qty(\frac{1}{\varepsilon_\infty}-\frac{1}{\varepsilon_r})
\end{align}
and $V_{p}$ in the local scattering approximation
\begin{align}
V_{p}(k_\parallel) = \frac{1}{\sqrt{k_\parallel^2+q_D^2}}\qty(1-\frac{q_D^2}{2(k_\parallel^2+q_D^2)}),
\end{align}
where $q_D$ is the inverse Debye screening length. The Green's function computed from the Dyson equation in Eq.~\ref{eq:Dyson} is solved self-consistently with the scattering functions in the self-consistent Born approximation (SCBA) scheme. The self consistent solution ensures current conservation along the spatial dimension, even when the spectral current density changes as it is influenced by the scattering.

\begin{figure}[h!]
\includegraphics[width=\columnwidth]{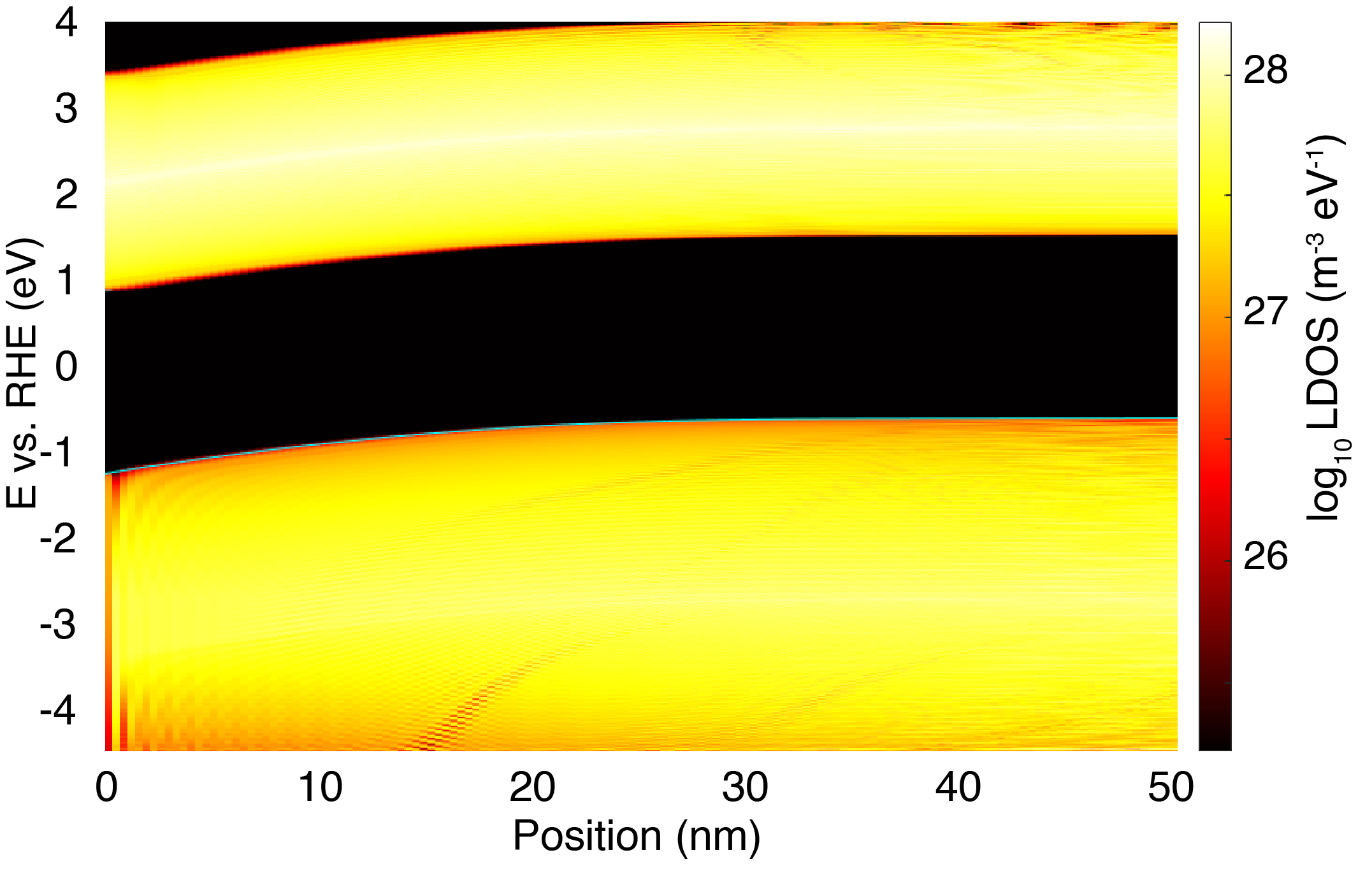}
\caption{\label{fig:DOS} Local density of states integrated over the transverse momentum at 1V vs. RHE. Due to the homogeneous structure of the device the band structure is also uniform with the exception of the band bending caused by the Schottky-like contact at the SEI.}
\end{figure}

\subsection{Boundary conditions}
Boundary conditions are required for both the electrostatic potential in Poisson's equation, and for the electrochemical potentials for both electrons and holes to find the boundary self energies. For the Poisson equation, at the SEI the potential of the conduction and valence band edges are fixed with respect to the RHE by using a Dirichlet boundary condition. At the rear contact, a Neumann condition is used instead to maintain zero electric field and the equilibrium carrier concentration. The boundary conditions for the Dyson equation are all Dirichlet conditions, fixing the electrochemical potentials and the carrier densities at the boundaries. They are defined by boundary self energies, which describe the interaction between the simulated device and electrical contacts at the boundaries. These boundary self energies depend on the electrochemical potential of the charge carriers in the contact reservoir and the Fermi-Dirac distribution at the operating temperature.\cite{jiang2008boundary,pal2008negf} For the simulation in this study, we do not consider any impurity states in the bandgap. The self energies for both the left (SEI) and right (metal contact) are given by
\begin{align}
\Sigma^{< (L,R)}_{b,k_\parallel} &= i f(\mu_{b}^{(L,R)}) \Gamma_{b,k_\parallel}^{(L,R)}\\
\label{eq:BoundarySE}
\Sigma^{> (L,R)}_{b,k_\parallel} &= -i(1-f(\mu_{b}^{(L,R)})) \Gamma_{b,k_\parallel}^{(L,R)}
\end{align}
where the superscripts $L$ and $R$ refer to the left and right contact, $f$ is the Fermi-Dirac distribution at electrochemical potentials $\mu_b$ and standard temperature and $\Gamma_{b,k_\parallel}$ are the broadening functions
\begin{align}
\Gamma_{b,k_\parallel}^{(L,R)} = i(\Sigma^{r (L,R)}_{b,k_\parallel}-\Sigma^{r (L,R) \dagger}_{b,k_\parallel}).
\end{align}
The electrochemical potentials at the rear contact are directly defined by the applied voltage vs. the reference electrode, $V_{ref}$, which is the reference voltage that is measured in experimental studies, giving a natural choice for a variable to control the external applied bias voltage. The electrochemical potentials at the electrolyte contact, however, require a more detailed description taking into account the selectivity of the contact and the kinetics of the chemical reaction.  Assuming a perfectly selective contact, for a p-type electrode the hole flux through the SEI is always zero. Therefore, the electrochemical potential of the holes in the electrolyte contact has no influence on the simulation. The electrochemical potential of the electrons on the other hand depends on the reaction potential of the chemical reaction driven by the electrode. In the case of p-type semiconductor electrode, the relevant reaction is the hydrogen evolution reaction
\begin{align}
\ce{2e- + 2H+ -> H2},
\end{align}
for which the reaction potential is by definition equal to the reference potential given by the RHE. The reaction kinetics relate the overpotential $\eta$  to the total current density flowing through the device. In this work, we use a simplified form of the Butler-Volmer equation which neglects the contribution of the reverse reaction
\begin{align}
    \label{eq:ButlerVolmer}
    J = j_0 e^{\frac{0.5 \eta}{k_B T}},
\end{align}
where the constant $j_0$ is the exchange current density at the surface of the electrode.
The total current density generated in the semiconductor is calculated directly from the Green's functions 
\begin{align}
\label{eq:currentDensity}
    J_i = \sum_b \sum_{k_{\parallel}}  \frac{e}{\hbar}t \int G_{b;i,i+1}^{<}(k_{\parallel},E) - G_{b;i,i+1}^{>}(k_{\parallel},E) dE
\end{align}
As the total current density is a conserved quantity, the value of $J$ is constant for all values of the spatial position index $i$. The kinetic relation in Eq.~\ref{eq:ButlerVolmer} couples the total current density to the overpotential $\eta$, which is defined as the difference between the electrochemical potential of the electrons at the SEI inside the semiconductor and the reaction potential of the half reaction driven by the electrode, in this case the hydrogen evolution reaction:
\begin{align}
    \eta = E_{2H^+/H_2}^0 - \mu_{c}^{(L)}.
\end{align}


Since the electrochemical potential at the SEI defines the boundary self-energy via Eq.~\ref{eq:BoundarySE} and therefore influences the Green's functions and the current density, a self-consistent solution scheme for the reaction kinetics is required. This is implemented by adjusting the Fermi level of the electron reservoir at the SEI boundary during the Poisson iteration until the current density given by Eq.~\ref{eq:currentDensity} also fulfills Eq.~\ref{eq:ButlerVolmer}. Correcting the Fermi level at the boundary simultaneously with the electrostatic potential offers considerable savings in computation time, as the self-consistent solution of the Green's functions, electric potential and the scattering self-energies already involves two nested iterative loops. Consequently, the Green's functions computed from the SCBA loop are self-consistently solved with both Poisson's equation for the electrostatic potential and the Butler-Volmer equation describing the kinetics of the hydrogen evolution reaction at the SEI.

In addition to the electrical current response, photoelectrochemical devices are commonly characterized by measuring the surface capacitance at the SEI. 
An ideal planar electrode with no surface states follows the well-established Mott-Schottky theory, where the interface capacitance at the SEI is measured as function of the bias voltage.~\cite{gelderman2007flat,hankin2019flat,sivula2021mott} The capacitance is commonly measured using electrochemical impedance spectroscopy (EIS), and should present a linear relationship between the inverse square of the capacitance and the bias voltage. As the simulation results give direct access to the carrier density distribution in the device at any given bias voltage, the Mott-Schottky response can be computed from the results as well. Unlike experimental setting where frequency domain measurements are needed to measure the capacitance, the simulation directly reveals the charge density. The carrier densities for both electrons and holes can be directly computed from the Green's functions

\begin{align}
    n_{i} = \sum_{k \parallel} \int  \frac{1}{\pi \Delta_z} G_{(c);i,i}^{<}(k_{\parallel},E) dE\\
    p_{i} = \sum_{k \parallel} \int  \frac{1}{\pi \Delta_z} G_{(v);i,i}^{>}(k_{\parallel},E) dE.
\end{align}
When the carrier densities are known, the surface capacitance as a function of the bias voltage can be calculated directly from the definition of differential capacitance
\begin{align}
    C(V_{ref})= \pdv{\rho_s}{V_{ref}},
\end{align}
where $\rho_s$ is the total excess charge density at the SEI, given by the sum of the excess hole and electron densities
\begin{align}
    \rho_s(V_{ref}) = \int_0^d p(d)-p(z)dz - \int_0^d n(d)-n(z)dz.
\end{align}

\section{\label{sec:Results}Results}

\begin{figure*}[ht]
\includegraphics[width=2\columnwidth]{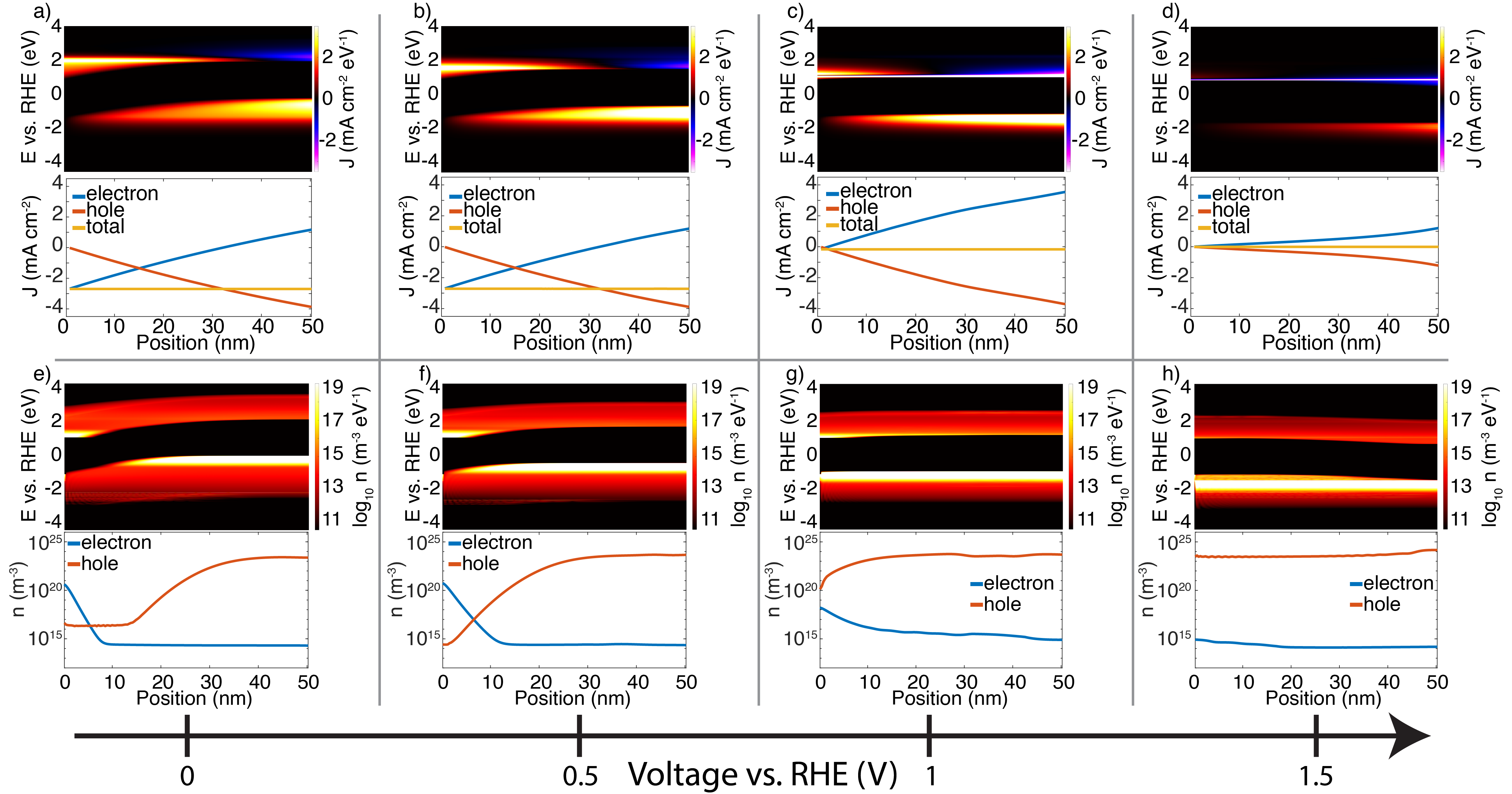}
\caption{\label{fig:denJ} \textbf{a)-d)} Color surface charts show the k-integrated, energy resolved current density $J$ as function of the applied voltage $V_{ref}$. Positive current is defined as positive charge flowing away from the SEI (left to right). The line charts show the same current densities, also integrated over energy. \textbf{e)-h)} Like a)-d), but for carrier density $n$. Larger bias voltages force the rear contact Fermi level to lower energy, eventually stopping the current flow at the SEI and all photogenerated carriers recombining at the rear contact.}
\end{figure*}

\begin{figure}[htb]
\includegraphics[width=\columnwidth]{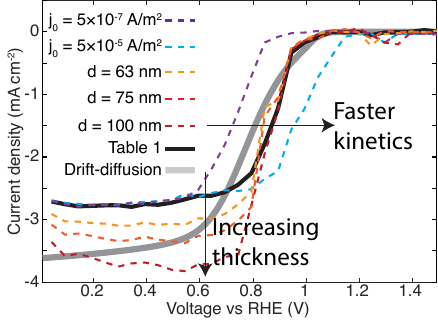}
\caption{\label{fig:IV} Total current as function of the rear contact voltage vs. RHE. The magnitude of the current increases towards more negative voltages as expected for a p-type device. Continuos line shows the i-V response with the parameters from Table~\ref{tb:material}. The dashed lines show the effect of either increasing or decreasing the exchange current density by one order of magnitude, or increasing the sample thickness to 100nm.}
\end{figure}

The simulation reproduces the typical I-V behaviour of a p-type photoelectrochemical cell as seen in Fig. \ref{fig:IV}. As the chemical reaction in a homogeneous PEC cell is driven by the minority carriers, in a p-type cell the hole current across the SEI is zero. In order for the electrons to have sufficient energy to drive the hydrogen evolution reaction, the rear contact Fermi level must be increased, which in terms of applied voltage means a more negative value measured vs. RHE. Since the device is a photocathode driving a reduction reaction, the reaction involves the transfer of electrons from the semiconductor into the electrolyte and the sign of the current density is negative. The simulated photocurrent onset potential is at \SI{1}{V} as seen in Fig. \ref{fig:IV}, which is just slightly below the theoretical minimum of \SI{1.23}{V} that would be required for unassisted water splitting. The required overpotential to drive the current depends on the kinetics of the hydrogen evolution reaction, which in the case of the presented model is described by the exchange current density $j_0$. Faster kinetics, meaning larger value of the exchange current density, result in lower overpotential $\eta$ needed for the reaction. Consequently, the the I-V curve is shifted towards more positive voltages, signifying that less external voltage is required to drive the water splitting reaction. This result is in line with previous studies using drift-diffusion models and experimental results showing that improved photocurrent onset potential can be achieved by utilizing more efficient catalysts that reduce the overpotential needed for the chemical reaction.~\cite{cendula2014calculation,hallstrom2021computational}
It is notable that faster kinetics do not increase the maximum current density. At high enough bias voltage the current density saturates at a constant value just above \SI{-2.5}{mA/cm^{2}} independent of the exchange current density, as the photocurrent density is ultimately limited by the photogeneration rate, which in turn is defined by the incident AM1.5G radiation and the efficiency of the photon-electron coupling given in Eq.~\ref{eq:photonelectronM}. With the optical losses caused by the glass window and the electrolyte, a perfect cell that would convert all photons above the \ce{Cu2O} bandgap of \SI{2.17}{eV} with unity external quantum efficiency would generate \SI{9}{mA/cm^2}. The external quantum efficiency (EQE) of the simulated device is therefore approximately 0.28. In the simulated case the saturated photocurrent is reached at approximately \SI{0.7}{V} vs. RHE, which would therefore be the optimal operating point for the simulated device. Given all the approximations used in the simulation, the simulated photocurrent density is a surprisingly good match to experimental results achieved with electrodeposited \ce{Cu2O} films~\cite{paracchino_synthesis_2012,cao2016facile,son2021design,qi2016cu2o}.

\begin{figure}
\includegraphics[width=\columnwidth]{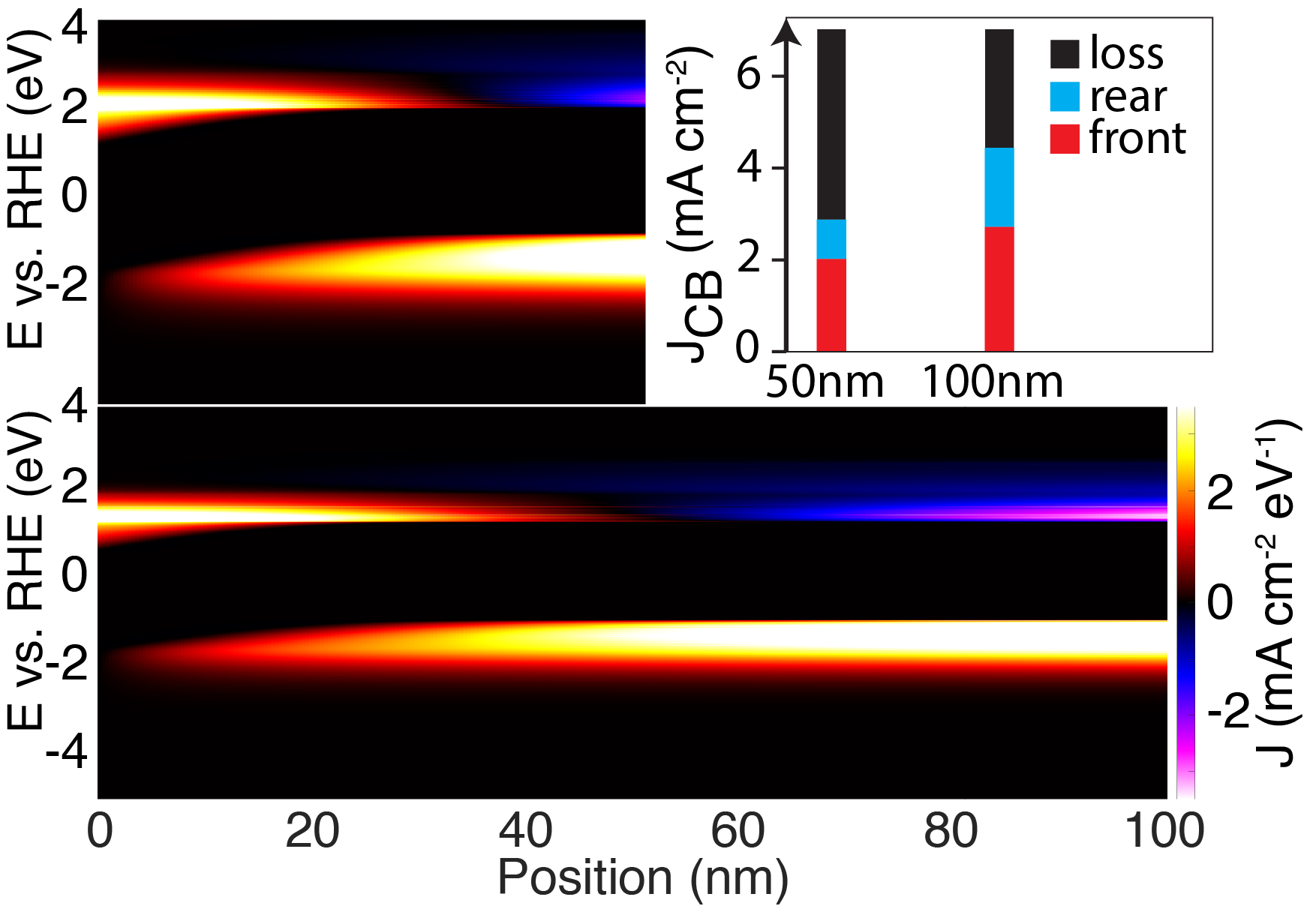}
\caption{\label{fig:thickness} Top left: spectral current density at 0.5V vs. RHE. Bottom: spectral current density at the same voltage, but with 100 nm thick electrode. Top right: fraction of conduction band current $J_{CB}$ at both contacts of the total useful incident photon flux.  Increasing the sample thickness to 100 nm absorbs more light, but the increase in useful photocurrent is reduced as majority of the electrons generated in the rear part of the device flow to the back surface and recombine. Loss includes non-absorbed photons and recombination in the semiconductor.}
\end{figure}

\begin{figure}
\includegraphics[width=\columnwidth]{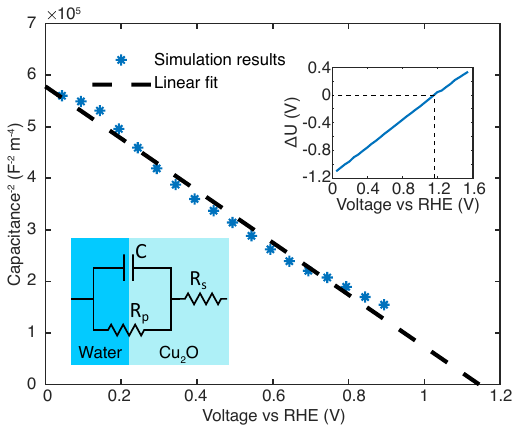}
\caption{\label{fig:CV} Simulated result of Mott-Schottky analysis. The slope is linear in the voltage range where the device generates current, with a slope corresponding to a doping density of \SI{4.65e17}{cm^{-3}}. The inset bottom left shows the commonly used Randles equivalent circuit, where C is the modeled capacitance. Top right inset shows the electrostatic potential drop in the \ce{Cu2O}, with zero potential at  \SI{1.15}{V}.}
\end{figure}

One of the advantages of the NEGF formalism is that all of the underlying physical quantities such as carrier and current densities can be directly extracted from the converged Green's functions. Since the device is a simple homogeneous semiconductor, the local density of states (LDOS) is mostly uniform over position. Figure~\ref{fig:DOS} shows the LDOS integrated over the transverse momentum $k_\parallel$ at \SI{0.5}{V} vs. RHE, with the typical band bending of the Schottky-like contact at the SEI clearly visible. As the metal contact at the backside is defined by applying a boundary condition enforcing zero electric field, the electrostatic potential and consequently the band edges are flat close to the rear contact, beyond the depletion region at the SEI. Inside the depletion region the depletion of the holes is accompanied with increased electron density, which leads to higher electrochemical potential for the electrons, eventually exceeding the hydrogen evolution reaction potential. Once sufficient overpotential is provided, the device starts generating current and driving the hydrogen evolution reaction. Figure~\ref{fig:denJ} shows the carrier density integrated over the transverse energy $k_\parallel$ at varying bias voltage values, clearly revealing the high electron density close to the SEI at \SI{0}{V} vs. RHE.

It is noteworthy that the electron current spectrum shown in Fig.~\ref{fig:denJ} is mostly located between 1-2 {V} vs. RHE, which corresponds to energy of ~\SI{0.5}{eV} above the conduction band edge at the SEI regardless of the bias voltage. The mean energy level of the current is significantly higher than the electrochemical potential at the SEI, signifying that only the high energy photoexited electrons participate in the chemical reaction. While the spectrum of the current density varies across the length of the device due to the scattering processes, the total current is always conserved. Figure.~\ref{fig:denJ} also shows that even at high bias voltages, carriers generated close to the rear contact are lost due to the electrons flowing to the metal contact instead of the SEI. This loss mechanism is well known in solar cells, and could be mitigated with the use of a heterojunction providing a selective barrier preventing the flow of electrons to the rear contact.~\cite{cavassilas2014modeling} However, the main loss factor reducing the EQE of the electrode is that a significant fraction of the incident light is not being absorbed in the thin \SI{50}{nm} \ce{Cu2O} layer. This can also be seen clearly in the line plots of the current density in Fig.~\ref{fig:denJ}, as both the electron and hole currents still increase in magnitude at the rear contact. In a region with no photon scattering, the currents in each individual band should be reducing due to spontaneous recombination, or be conserved if the carrier concentrations are in equilibrium.  As seen in Fig.~\ref{fig:thickness}, increasing the thickness to \SI{100}{nm} helps to absorb more of the incident photons, but not all of the gained current is converted to useful photocurrent. While increasing the thickness does increase the useful photocurrent, it also increases the fraction of the photogenerated electrons that are lost to recombination at the rear contact, leading to diminishing returns. The current spectrum of the \SI{100}{nm} electrode in Fig.~\ref{fig:thickness} shows that only a small fraction of the electrons generated in the region beyond \SI{50}{nm} flow to the SEI.

While the current spectrum mostly varies in magnitude, staying at the same energy at all voltages, the spectral carrier density follows the band edge shape with highest electron concentration found in the depletion region at the SEI. Increasing the voltage reduces the size of the potential well that is formed at the SEI due to the band bending, and eventually at \SI{1}{V} the flat band potential is reached, and the carrier densities are constant across the device. At the rear contact the hole density is always equal to the doping density at \SI{5e17}{cm^{-3}}, and the electron density depends on the photogeneration rate of the carriers. At higher voltages the excess electron concentration and hole depletion at the SEI is reduced, eventually resulting in the hole density across the semiconductor being in equilibrium with the rear contact and zero net current. While electron-hole pairs are still being generated by the incoming photons, all carriers flow to the rear contact and their energy is lost to recombination.

The result of the simulated Mott-Schottky response in Fig.~\ref{fig:CV} shows the expected linear relation between the inverse square of the surface capacitance and the applied voltage vs RHE. Using  the analytic model based on the full depletion approximation the effective doping density of the semiconductor can be estimated 
\begin{align}
    N_a \approx \frac{1}{\abs{k}}\frac{2}{\varepsilon_0\varepsilon_r q},
\end{align}
where $k$ is the slope from the linear fit in Fig.~\ref{fig:CV}. The fit gives a doping density of \SI{4.65e17}{cm^{-3}}, which is an excellent match to the simulation input value of \SI{5.0e17}{cm^{-3}}. Furthermore, the value for the flat band voltage found by extrapolating the fit in the Mott-Schottky plot at \SI{1.15}{V} vs. RHE matches the actual zero-potential voltage, shown in the inset of Fig.~\ref{fig:CV}, indicating that the relation between the surface charge at the SEI and the bias voltage is reproduced accurately by the simulation.

\section{Conclusion}
We have demonstrated that a quantum transport model based on NEGF formalism can offer a detailed picture of the carrier dynamics in the semiconductor electrode of a PEC cell. Even though numerous simplifying assumptions were used to reduce the complexity of the model and the computational cost, the simulation reproduces the correct current-voltage and Mott-Schottky behaviour that are experimentally observed in PEC systems. Compared to commonly used drift-diffusion models, the NEGF simulation offers considerable advantages as it requires less empirical parameters, and provides an energy resolved solution. The results shown in this work provide new insight into the carrier dynamics in photoelectrochemistry and offer a way to accurately model the influence of nanoscale design choices on the efficiency of the electrodes.

\section{Acknowledgment}
The authors acknowledge the financial support from the Academy of Finland project 329406 and the Photonics Research and Innovation (PREIN) flagship program, decision number 320167. L.H. acknowledges funding from the Aalto ELEC doctoral school, and Walter Ahlström foundation. Finally, we acknowledge the computational resources provided by the Aalto Science-IT project.
\clearpage

\bibliography{refs}

\end{document}